\documentclass[12pt]{article}
\usepackage{latexsym,graphicx,multirow}
\usepackage{float}
\usepackage{amssymb}
\usepackage{amsmath}
\usepackage{amscd}
\usepackage{amsthm}
\usepackage[left=2cm,top=2.5cm,right=2.5cm,bottom=1.5cm]{geometry}
\usepackage{hyperref}
\usepackage{epstopdf}
\usepackage{cite}

\begin{document}
	
 \begin{center}
 \large{\bf{RHDE models in FRW Universe with two IR cut-offs with redshift parametrization}} \\
\vspace{10mm}
 \normalsize{ Archana Dixit$^1$, Vinod Kumar Bhardwaj $^2$ Anirudh Pradhan$^3$  }\\
\vspace{5mm}
 \normalsize{$^{1,2,3}$Department of Mathematics, Institute of Applied Sciences and Humanities, GLA University, Mathura,\\
			Uttar Pradesh-281 406, India} \\
\vspace{2mm}
$^1$E-mail: archana.dixit@gla.ac.in.\\
$^2$E-mail: dr.vinodbhardwaj@gmail.com, vinod.bharadwaj@gla.ac.in\\
$^3$E-mail: pradhan.anirudh@gmail.com\\

\vspace{10mm}
	
		%\date{}
		%\maketitle
\end{center}

 \begin{abstract}
In this manuscript, we have researched the cosmic expansion phenomenon in flat FRW Universe through the interaction of the recently proposed 
 R$\grave{e}$nyi holographic dark energy (RHDE). For this reason, we assumed Hubble (H) and Granda--Oliveros (GO) horizons as IR cut-off 
 in the framework of $f(R, T)$ gravity. With this choice for IR cut-off, we can obtain some important cosmological quantities such as the 
 equation of state $\omega_{_T}$, energy density $\rho_{_T}$, density parameter $\Omega_{_T}$, and pressure $p_{_T}$, which are the function 
 of the redshift  $z$. It is observed that in both IR cut-offs the EoS parameter displays quintom-like behaviour for three different values of 
 $\delta$. Here we plot these parameters versus redshift $z$ and discuss the consistency of the recent findings.  Next, we explore the 
 $\omega_{_T}$-$\omega_{_T}^{'}$ plane and the stability analysis of the dark energy model by a perturbation method. Our findings demonstrate 
 that the Universe is an accelerating model of rapid growth that is explained by quintom like behaviour. Hence the feasibility of the RHDE model 
 with Hubble and GO cut-off is supported by our model. The results indicate that the IR cut-offs play a significant role in the understanding of 
 the dynamics of the universe. 

\end{abstract}
 
 
%%%%%%%%%%%%%%%%%%%%%%%%%%%%%%%%%%%%%%%%%%%%%%%%%%%%%% Section 1 %%%%%%%%%%%%%%%%%%%%%%%%%%%%%%%%%%%%%%%%%%%%%%%%%%%%%%%%%%%%%%
 \section{Introduction}

Observational information received by (SNIa) \cite{ref1}-\cite{ref2}, large scale structures (LSS) \cite{ref3}-\cite{ref5} and cosmic 
microwave background (CMB), anisotropies \cite{ref6}-\cite{ref7} confirmed the present accelerated expansion of the universe. The dark 
energy is assumed as a responsible candidate for this present scenario of accelerated Universe \cite {ref8}-\cite {ref10}. The character 
of DE is unknown, and mysterious. The easiest choice for DE is cosmological constant with positive energy density and negative pressure.\\

The cosmological constant faces several challenges, such as the issue of fine tuning, and the problem of coincidence \cite{ref8}. One 
reasonable way of relieving the question of cosmic coincidence is to assume that dark matter (DM) and dark energy (DE) interact. For the complex 
DE scenario, there are different alternate theories suggested by observing the accelerating universe: (a) the scalar-field models of DE 
including quintessence \cite{ref11}, tachyon \cite{ref12}, k-essence \cite{ref13}, (b) the interacting DE models including chaplygin 
gas \cite{ref14}, polytropic gas \cite{ref15}-\cite{ref16}, phantom \cite{ref17} and holographic dark energy (HDE) \cite {ref18}-\cite {ref19}. 
On the other hand, the nature of DE can be investigated on the basis of certain principles of quantum gravity, which would be the yield to 
the (HDE) model\cite {ref20}-\cite {ref24}.\\

Another approach to solving the dark energy problem is to connected the certain aspects of the theory of quantum gravity, known as the holographic 
principle \cite{ref25}-\cite{ref28}. Several authors are focusing on the numerous cosmological implications of new and modified 
HDE models \cite{ref29}-\cite{ref33}. Another significant factor of this model is the IR cut-off. A lot of HDE models are discussed in 
the literature, depending on the IR cut-off.\\

Many authors are working on the different IR cut-offs like: Sharma and Dubey \cite{ref34} worked on the interacting R$\grave{e}$nyi holographic dark energy 
with parametrization on the interaction term and studied the graphical behaviour of cosmological parameters. Vipin et al. \cite{ref35} have also 
observed the growth rate of perturbations by using hierarchy. The anisotropic and spatially homogeneous Bianchi type-$VI_{0}$  (RHDE) models 
of general relativity discussed in \cite{ref36}. In this model authors assumed both (H) and (GO) horizons as IR cut-off and obtained cosmological parameters. 
In the same context, Tayeb {\it et al.} \cite{ref37} analysed the anisotropic R$\grave{e}$nyi holographic dark energy models in flat space-time. 
Qolibikloo and Ghodsib \cite{ref38} investigate the Rényi entropy and inequalities under the phase transition. The expansion and growth data were combined by 
Akhlaghi \cite{ref39} to investigate the ability of the three most popular HDE models, namely the Ricci scale, future event horizon, and Granda--Oliveros IR cut-offs.\\

In this sequence, Ghaffari \cite{ref40} examined the cosmological models of HDE in a DGP braneworld with GO cut-off. In recent year many 
entropy formalisms  has been used and explore the cosmological models. Some new HDE models are being developed, such as the RHDE model,
\cite{ref41}, Sharma--Mittal HDE (SMHDE) \cite{ref42} and Tsallis HDE (THDE). Among these models RHDE is more stable model which is based on non-interactions 
between cosmic region \cite{ref43}-\cite{ref44}. Younas {\it et al.} \cite{ref45} have investigated entropies of R$\grave{e}$nyi, Tsallis and 
Sharma--Mittal in flat FRW Universe within Chern--Simons modified gravity. HDE model was conjectured as IR cut-off with Benkenstein entropy and Hubble 
horizon that does not provide an appropriate explanation for the level of a flat FRW universe \cite{ref46}-\cite{ref48}. Recently, a new HDE model was 
proposed and explored by changing the standard HDE as $S_{\delta}= \gamma A^{\delta}_{r}$, where $\gamma$ is an unspecified constant $A_{r}= 4\pi L^{2}$ 
represents the area of the horizon and $\delta$ is a non-additivity parameter, called Tsallis holographic dark energy (THDE) \cite{ref44,ref49}. Another 
possibility for DE showed up when Cohen {\it et al.} \cite{ref50} applied some speculation on the mutual connection between UV $(\Lambda)$ and IR (L) cut-offs 
and the entropy of framework, expressing as $\rho_{_\Lambda}\varpropto{\frac{S}{L^{4}}} $. where $\rho_{_\Lambda}$ is the vacuum energy density \cite{ref50}. 
Bekenstein and Hawking \cite{ref51,ref52} studied  the thermodynamics of the black hole. It is indicated that the Bekenstein--Hawking 
entropy bound $S_{_BH}\sim M_{p}^{2}L^{2}$, scales as the region $A_{r}\backsim L^{2}$ rather than the volume $V\backsim L^{3}$ and  $M_{p}$ 
is the reduced Plank mass ($8\pi G = 1/M_{p}^{2} = 1$). On the other hand, the Bekenstein entropy bound $S_{_B}$ is $EL$ for the case where 
$E = \rho_{_\Lambda} L^{3}$ is the energy, and $L$ is the IR cut-off. For $\rho_{_\Lambda} \leq M_{p}^{2} L^{-2}$ by using $S_{B} < S_{_BH}$. 
Here $C^{2}_{1}$ is a numerical constant, given the HDE:

\begin{equation}
\label{1}
\rho_{d}= 3C^{2}_{1}M_{p}^{2}L^{-2}.
\end{equation} 

Observational information, which is acquired by restricting the HDE model, clearly demonstrate $C_{1} = 0.818+_{-0.097}^{0.113}$ \cite{ref53} 
and $C_{1} =0.815+_{-0.039}^{0.179}$ \cite{ref54} for flat and non-flat space time separately. Several research on the HDE model and its 
features are reported in \cite{ref55}-\cite{ref58}. The two latest entropies R$\grave{e}$nyi and Tsallis \cite{ref59}-\cite{ref62} are commonly used 
to study various gravitational and cosmological phenomena \cite{ref63}-\cite{ref71}. HDE density derivation is based on the relationship of 
entropy-area  $S_{BH}= A_{r}/4$, where $A_{r}$ is the area of black hole horizon. Subsequently, by changing the entropy relation, one can locate 
new type of HDE. It is surprising that the form of HDE and gravity model equations can be generalized by using a generalized method of entropy-area 
relationship. So, we have a generalized Friedmann equation to define evolution of the universe. This idea motivates us to look through 
R$\grave{e}$nyi entropy into the constructive accelerating phases of the Universe. The entropy in RHDE was discussed in Refs.\cite{ref65}-\cite{ref67}. 

\begin{equation}
\label{2}
S_{_R}=\frac{1}{\delta}ln(1+\delta S_{_T}).
\end{equation} 
Here  $\delta$ is a constant and  $S_{_T}$ is Tsallis entropy. Bekenstein and Tsallis is equal \cite{ref65,ref66,ref67,ref71} then 
Eq. (\ref{2}) becomes:
\begin{equation}
\label{3}
S_{_R}=\frac{1}{\delta}ln(1+\delta \frac{A_{r}}{4}),
\end{equation} 
if $\delta$ tends to zero, the R$\grave{e}$nyi entropy reduces to $A_{r}/4$. During the current work, we are considering HDE by applying the R$\grave{e}$nyi entropy. 
In this direction a brief survey of the theories of modified gravity as the new participation explored a cosmological reconstruction. 
Harko {\it et al.} \cite{ref72} developed new generalized theory known as $F(R,T)$ gravity.\\ 

Several cosmologist \cite{ref73}-\cite{ref75} have studied $f(R, T)$ gravity in distinct context. The main aim of our proposed model is to consider 
$f(R, T)$ model with R$\grave{e}$nyi HDE by assuming two IR cut-offs. The work in this manuscript is configured as: The field equations of $f(R, T)$ 
gravity is set out in Sect.$2$. R$\grave{e}$nyi HDE models are considered in Sect.$3$. R$\grave{e}$nyi HDE model with Hubble horizon cut-off are discussed in 
Sect.$3.1$. R$\grave{e}$nyi HDE model with GO horizon cut-off is examined in Sect.$3.2$. $\omega_{_T}-\omega_{_T}^{'}$ plane is discussed in Sect.$4$. The stability of the model is discussed in Sect.$5$ and the outcomes are summarized with conclusions that are discussed in Sect.$6$.

%%%%%%%%%%%%%%%%%%%%%%%%%%%%%%%%%%%%%%%%%%%%%%%%%%%%%%% Section 2 %%%%%%%%%%%%%%%%%%%%%%%%%%%%%%%%%%%%%%%%%%%%%%%%%%%%%%%%%%%%%%%%

\section {Basic field equations of f(R,T) gravity} 

We assume that the behaviour for the modified gravity theories takes the form:
\begin{equation}
\label{4}
{\bf \mathbb{S}}=\frac{1}{16\pi}\int f(R,T)\sqrt{-g} d^{4}x +\int L_{m}\sqrt{-g}d^{4} x.
\end{equation}
Here $f(R, T)$ is an arbitrary function of the trace ($T$) and Ricci scalar ($R$). Here $L_{m}$ is the matter Lagrangian density.\\

The stress-energy tensor of matter is defined as \cite{ref76}
\begin{equation}
\label{5}
T_{ij}= -\frac{2}{\sqrt{-g}}\frac{\delta (\sqrt{-g} L_{m})}{\delta g^{ij}}.
\end{equation}

The field equations of the $F(R, T)$ model are obtained as
 \begin{equation}
 \label{6} 
  2 F_{_R}(R,T)R_{ij}-F(R,T)g_{ij}+2(g_{ij}\square-\nabla_{i}\nabla_{j})f_{_R}(R,T)= 16\pi T_{ij}-2F_{_T}(R,T)T_{ij}-2F_{_T}(R,T)\ominus_{ij}.
 \end{equation}

However, the stress-energy tensor for the current work is considered as
 \begin{equation}
 \label{7}
 T_{ij}=  -pg_{ij}+(p+\rho) u_{i}u_{j}.
 \end{equation}
Also, the matter Lagrangian can be assumed as $L_{m}=-p$. The conditions  $u_{i}u^{i}=1$ and  $u^{i}\nabla_{j}u_{i}=0$ satisfy the 
four-velocity. In our model, we assumed the  particular case of $f(R,T)$ expressing by the function $f(R,T)= 2f(T)+R $, where f(T) is an 
arbitrary function of the stress-energy tensor of matter. The gravitational field equations follows by Eq. (\ref{6}) written as
\begin{equation}
\label{8}
2R_{ij}-Rg_{ij}=16\pi T_{ij}+4 \dot f(T)T_{ij}+2[2p\dot f(T)+f(T)]g_{ij},
\end{equation}
where the dot indicates the derivative is related to the argument.
For the dust filled universe ($p = 0$), the gravitational field equations are given by
\begin{equation}
\label{9}
2R_{ij}-Rg_{ij}=16\pi T_{ij}+4\dot f(T)T_{ij}+2f(T)g_{ij}.
\end{equation}
These field equations are suggested in \cite{ref77}. We can obtain a cosmological model by considering the dust Universe, by select the 
function  $f(T) = \xi T$, $\xi \rightarrow$ constant. \\

The metric of a flat FRW Universe is considered as:
\begin{equation}
\label{10}
ds^{2}=dt^{2}-a^{2}\left(dx^{2}+dy^{2}+dz^{2}\right).
\end{equation}
The gravitational field equations are written as,
\begin{equation}
\label{11}
3\left(\frac{\dot a^{2}}{a^{2}}\right)= (8\pi+3\xi)\rho_{T},
\end{equation}
\begin{equation}
\label{12}
 2\frac{\ddot a}{a}+\frac{\dot a^{2}}{a^{2}}= \xi\rho_{T}.
\end{equation}
This $f(R, T)$ gravitational model is equal to efficient cosmological constant $\Lambda_{eff}\varpropto H^{2}$, where 
$ H = \frac{\dot a}{a}$ is the Hubble function \cite{ref77}. Field equations are reduced in terms of $H$,
\begin{equation}
\label{13}
2\dot H+3\frac{8\pi+2\xi}{8\pi+3\xi}H^{2}=0.
\end{equation}
 The general solution of the above Eq. (\ref{13}) obtained as
\begin{equation}
\label{14}
H(t)=\frac{2}{3 t}\frac{(8\pi+3\xi)}{(8\pi+2\xi)} = \frac{2 \beta}{3 t},
\end{equation}
where $ {\bf  a(t)= t^{\frac{2}{3}\beta}}$ is the scale factor where $\beta =\frac{(8\pi+3\xi)}{(8\pi+2\xi)}$.\\

Here we are using different expression for $H$, like future event Hubble horizon as the infrared cut-off, Ref.\cite{ref78} and GO infrared cut-off. 
By the conservation equation we obtained as
\begin{equation}
\label{15}
\frac{\partial \rho_{_T}}{\partial t}+3H(\rho_{_T}+p_{_T})=0.
\end{equation}
Here $\rho_{_T}$ is the R$\grave{e}$nyi holographic energy density, solving Eq.(\ref{15})
the EoS parameter $\omega_{_T}$ can be rewrite as:
\begin{equation}
 \label{16}
 \omega_{T}= -1- \frac{1}{3H}\left(\frac{\dot \rho_{_T}}{\rho_{_T}}\right).
 \end{equation}
 
 %%%%%%%%%%%%%%%%%%%%%%%%%%%%%%%%%%%%%%%%%%%%%%%%%%%%%%% Section 3 %%%%%%%%%%%%%%%%%%%%%%%%%%%%%%%%%%%%%%%%%%%%%%%%%%%%%%%%%%%%%
 \section{ R$\grave{e}$nyi HDE model  }
 
We have taken a system with $m$, states with $P_{i}$ probability distribution  and satisfies the condition $\sum_{i=1}^{m} P_{i}=1$. 
R$\grave{e}$nyi and Tsallis entropy are well known parameter of generalized entropy \cite{ref49}.

\begin{equation}
\label{17}
S_{_R}=\frac{1}{\delta}ln\sum_{i}^{m}P_{i}^{1-\delta}~~~~~ S_{_T}=\frac{1}{\delta}ln\sum_{i=1}^{m}P_{i}^{1-\delta}- P_{i},
\end{equation}
where $\delta=1-U $ and $U$ is a real parameter. Now using the above equations we get the relation,

\begin{equation}
\label{18}
S_{_R}=\frac{1}{\delta}ln(1+\delta S_{T}).
\end{equation}

In Eq. (\ref{18}) the Bekenstein entropy is 
 $S_{T} = \frac{A_{r}}{4}$, ~ and $A_{r}= 4\pi L^{2}$.
This gives the R$\grave{e}$nyi entropy of the system as $S_{R}=\frac{1}{\delta}ln(1+\delta\pi L^{2})$ \cite{ref60}.\\

In this section  we take the assumption $\rho_{_T}dV \varpropto Tds$ \cite{ref69} then we can get the R$\grave{e}$nyi HDE density in IR cut-off written as 

\begin{equation}
\label{19}
\rho_{_T}=\frac{3 C^2_{1}}{8 \pi L^2}\frac{1}{\left(1+\pi \delta L^2\right)}.
\end{equation}

In this case, $C^{2}_{1}$ is the numerical constant, $V$ is the  volume, and $T$ is the  temperature of the system. We have used 
$T=\frac{H}{2\pi}$ and $A = \frac{4\pi}{H^{2}}= 4\pi(\frac{3V}{4\pi})^{2/3}$, relationships valid for the flat FRW Universe. It is clear 
that we have $\rho_{_T}=\frac{3 C^{2}_{1} H^{2}}{8\pi}$ is a complete agreement with OHDE \cite{ref21,ref46,ref47} in the absence of $\delta$. 
In this section, we use the R$\grave{e}$nyi entropy by considering the Bekenstein entropy as the Tsallis entropy \cite{ref63}-\cite{ref69}. \\

Finally, we are proposing a new holographic dark energy model, which is RHDE. Next we have examined the well-known cosmological parameters of $H$ 
and GO IR cut-off in Sect.3.1 $\&$  Sect.3.2. 

%%%%%%%%%%%%%%%%%%%%%%%%%%%%%%%%%%%%%%%%%%%%%%%%%%%%%%%%%%%%%% SubSection 3.1 %%%%%%%%%%%%%%%%%%%%%%%%%%%%%%%%%%%%%%%%%%%%

\subsection{Model-I: R$\grave{e}$nyi HDE model with hubble horizon cut-off with redshift parameterization}

Hubble horizon is the simplest choice with the Hubble length $L = H^{-1}$ \cite{ref48}. According to the holographic principle the energy 
density of DE is proportional to the square of the Hubble parameter i.e., $\rho_{T} \propto H^{2}$ \cite{ref39}. The principle states that 
the given choice can solve the fine tuning problem. \\

Here the R$\grave{e}$nyi energy density is obtained as

\begin{equation}
\label{20}
\rho_{_T}=\frac{3 C^2_{1}H^{2}}{8 \pi}\left(1+ \frac{\pi \delta}{H^{2}}\right)^{-1}
\end{equation} 
Utilizing Hubble horizon as a possibility for IR cut-off
  i.e., $ L=\frac{1}{H} $ and $8\pi=1$. In this analysis we use the $a=\frac{a_{0}}{(1+z)}$  redshift parameterization,
\begin{equation}
\label{21}
\rho_{_T}=\frac{2 \beta ^4 C^2_{1}}{12 \pi  \beta ^2 \left(\frac{a_0}{z+1}\right){}^{3/\beta }+27 \pi ^2 \delta 
\left(\frac{a_0}{z+1}\right){}^{6/\beta }}.
\end{equation}

By using  Eqs.(\ref{16}) and (\ref{21}), we get EoS parameter defined as,
\begin{equation}
\label{22}
\omega_{_T}=\frac{-9 \pi  \beta  \delta  \left(\frac{a_0}{z+1}\right){}^{3/\beta }+18 \pi  \delta 
\left(\frac{a_0}{z+1}\right){}^{3/\beta }-4 \beta ^3+4 \beta ^2}{9 \pi  \beta  \delta  \left(\frac{a_0}{z+1}\right){}^{3/\beta }+4 \beta ^3}.
\end{equation}
By using Eqs.(\ref{21}) and (\ref{22}) we get pressure as,
\begin{equation}
\label{23}
p_{_T}=-\frac{2 \beta ^3 c^2 \left(9 \pi  \beta  \delta  \left(\frac{a_0}{z+1}\right){}^{3/\beta }-18 \pi  \delta 
\left(\frac{a_0}{z+1}\right){}^{3/\beta }+4 \beta ^3-4 \beta ^2\right)}{3 \pi  \left(4 \beta ^2 \left(\frac{a_0}{z+1}\right){}^{\frac{3}
{2 \beta }}+9 \pi  \delta  \left(\frac{a_0}{z+1}\right){}^{\frac{9}{2 \beta }}\right){}^2}.
\end{equation}
Using Eqs.(\ref{14}) and (\ref{21}), we obtain density parameter as
\begin{equation}
\label{24}
\Omega_{_T}=\frac{\beta ^2 C^2_{1}}{18 \pi ^2 \delta \left(\frac{a_0}{z+1}\right){}^{3/\beta }+8 \pi \beta ^2}.
\end{equation}

The behaviour of cosmological parameters such as energy density, Eos parameter, pressure and density parameter are shown in this section, where the 
density parameter  $\Omega_{T}$ is defined as the ratio between the actual (or observed) density $\rho_{T}$ and the critical density $\rho _{c}$ of the 
Friedmann Universe. Here $\rho {c}=\frac{3H^{2}}{8\pi}$, the relation between the actual density and the critical density determines the overall geometry 
of the Universe; when they are equal, the geometry of the Universe is flat (Euclidean) of the total RHDE in IR cut-off.

%%%%%%%%%%%%%%%%%%%%%%%%%%%%%%%%%%%%%%%%%%%%%%%%%%%%%% SubSEction 3.2 %%%%%%%%%%%%%%%%%%%%%%%%%%%%%%%%%%%%%%%%%%%%%%

\subsection{Model-II: R$\grave{e}$nyi HDE model with Granda--Oliveros (GO) horizon cut-off with redshift parameterization}

Granda and Oliveros first \cite{ref79,ref80} introduced the cut-off $ L=(k H^2+m \dot{H})^{-1/2}$ and attempt to resolve the well 
known cosmological problems like cosmic coincidence and stability. In this sequence, the acceleration combined the Hubble parameter 
together with its time derivative. On substituting GO cut-off in Eq. (\ref{14}) and taking $(8\pi = 1)$, we get 
$ L=\frac{1}{\frac{4}{9} \beta ^2 k \left(\frac{a_0}{z+1}\right){}^{-3/\beta }-\frac{2}{3} \beta  m \left(\frac{a_0}{z+1}\right){}^{-3/\beta }}.$ \\

By using Eq. (\ref{20}), we get energy density as,
\begin{equation}
\label{25}
\rho_{_T}=\frac{2 \beta ^4 c^2 \left(\frac{a_0}{z+1}\right){}^{-6/\beta } (3 m-2 \beta  k)^4}{27 \pi  \left(81 \pi  \delta  
\left(\frac{a_0}{z+1}\right){}^{6/\beta }+16 \beta ^4 k^2-48 \beta ^3 k m+36 \beta ^2 m^2\right)}.
\end{equation}
By using Eqs. (\ref{16}) and (\ref{25}), we get EoS parameter defined as
\begin{equation}
\label{26}
\omega_{_T}=\frac{-81 \pi  (\beta -4) \delta  \left(\frac{a_0}{z+1}\right){}^{6/\beta }-16 (\beta -2) \beta ^4 k^2+48 (\beta -2) 
\beta ^3 k m-36 (\beta -2) \beta ^2 m^2}{81 \pi  \beta  \delta  \left(\frac{a_0}{z+1}\right){}^{6/\beta }+ 
16 \beta ^5 k^2-48 \beta ^4 k m+36 \beta ^3 m^2}.
\end{equation}

By using Eqs.(\ref{25}) and (\ref{26}) we obtained pressure as
\begin{equation}
\label{27}
p_{_T}=\frac{2 \beta ^3 c^2 \left(\frac{a_0}{z+1}\right){}^{-6/\beta } (3 m-2 \beta  k)^4 \left(-81 \pi  (\beta -4) \delta  
\left(\frac{a_0}{z+1}\right){}^{6/\beta }-4 (\beta -2) \beta ^2 (3 m-2 \beta  k)^2\right)}
{27 \pi \left(81 \pi \delta \left(\frac{a_0}{z+1}\right){}^{6/\beta }+4 \beta ^2 (3 m-2 \beta  k)^2\right){}^2}.
\end{equation}
Using Eqs.(\ref{14}) and (\ref{25}) we get density parameter as

\begin{equation}
\label{28}
\Omega_{_T}=\frac{\beta ^2 c^2 \left(\frac{a_0}{z+1}\right){}^{-3/\beta } (3 m-2 \beta  k)^4}{18 \pi 
\left(81 \pi  \delta  \left(\frac{a_0}{z+1}\right){}^{6/\beta }+16 \beta ^4 k^2-48 \beta ^3 k m+36 \beta ^2 m^2\right)}.
\end{equation}

Here we also calculate the well known cosmological parameters like energy density, Eos Parameter, pressure and  density parameter 
$\Omega_{_T}$  of the Friedmann Universe.  We noticed that the geometry of the universe is flat (Euclidean) of the total RHDE in GO cut-off.

%%%%%%%%%%%%%%%%%%%%%%%%%%%%%%%%%%%%%%%%%%%%%%%%%%%%%%%% Figure 1(a) & 1(b) %%%%%%%%%%%%%%%%%%%%%%%%%%%%%%%%%%%%%%%%%%%
	
	\begin{figure}[H]
		
		\centering
		(a)\includegraphics[width=7cm,height=7cm,angle=0]{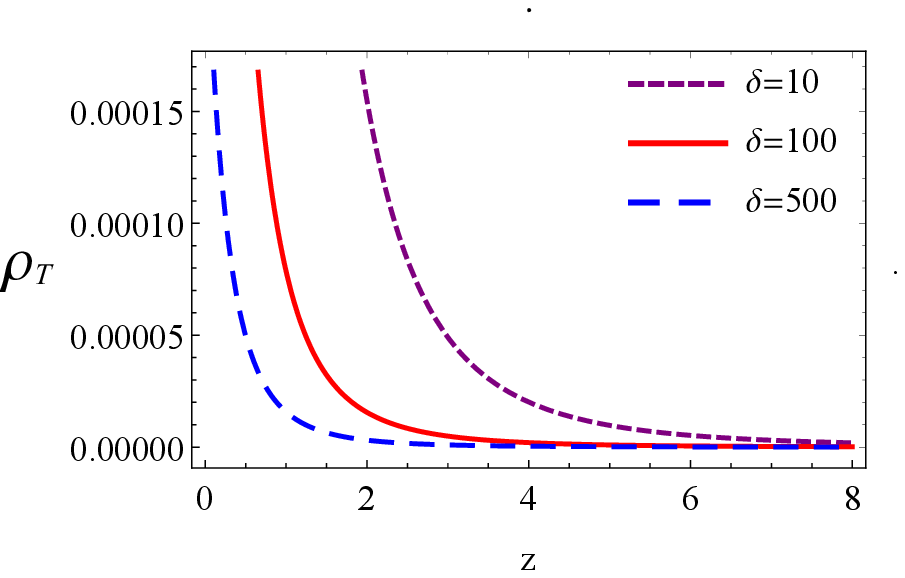}
		(b)\includegraphics[width=7cm,height=7cm,angle=0]{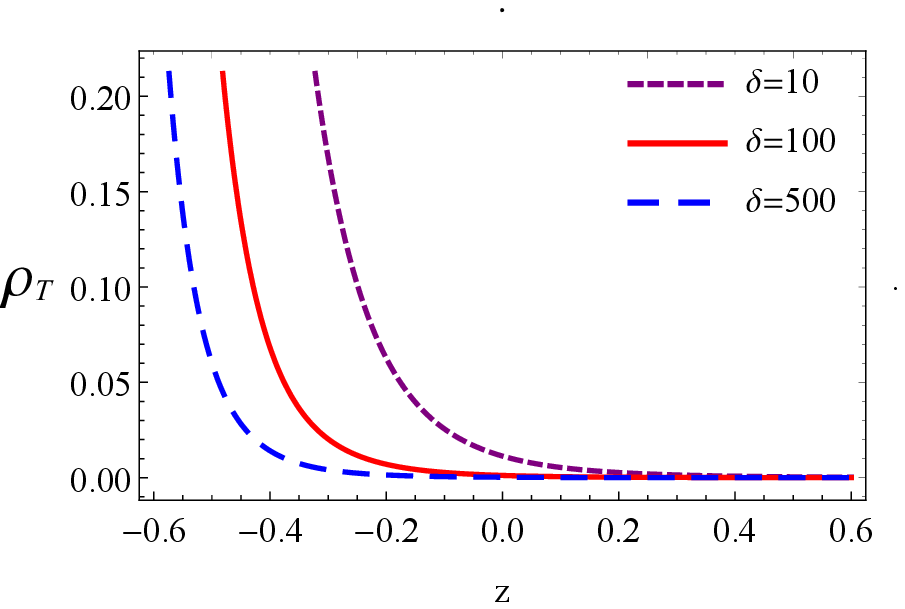}
		\caption{a. Plot of $ \rho_{_T} $ versus  $ z $  with (H hoz.) cut-off 
			b. Plot of  $ \rho_{_T} $ versus $z$ with  (GO hoz.) cut-off}	
	   \end{figure}
	%%%%%%%%%%%%%%%%%%%%%%%%%%%%%%%%%%%%%%%%%%%%%%%%%%%%%%%%%%%%%%%%%%%%%%%%%%%%%%%%%%%%%%%%%%%%%%%%%%%%%%%%%
	
  Figure $1a, b$ show the energy density of RHDE with Hubble horizon and GO horizon cut-off versus redshift $z$ respectively. All the 
  trajectories of $\rho_{_T}$ indicate the positive behaviour throughout the evolution of the universe for various estimations of  $\delta$. 
  Likewise it can be seen that $\rho_{_T}$ is a positive decrease function and decreases, more sharply as $\delta$ increases.
  We observed RHDE is the decreasing function of redshift $z$ in both IR cut-offs and we also found that in high redshift 
  $\rho_{_T}$ tends to zero.
   
   %%%%%%%%%%%%%%%%%%%%%%%%%%%%%%%%%%%%%%%%%%%%%% Figure 2(a) & 2(b) %%%%%%%%%%%%%%%%%%%%%%%%%%%%%%%%%%%%%%%%%%%%%
   
   	\begin{figure}[H]
   	
   \centering
   	(a)\includegraphics[width=7cm,height=7cm,angle=0]{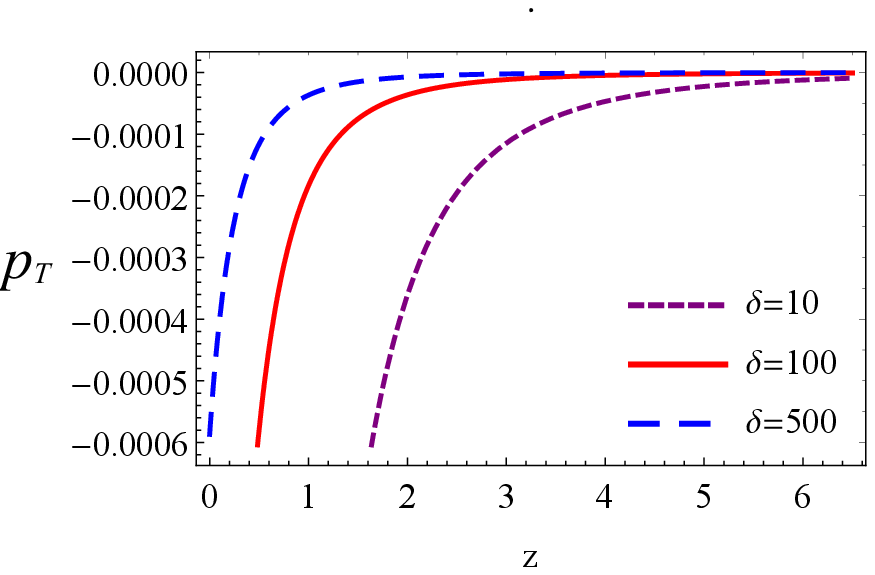}
   	(b)\includegraphics[width=7cm,height=7cm,angle=0]{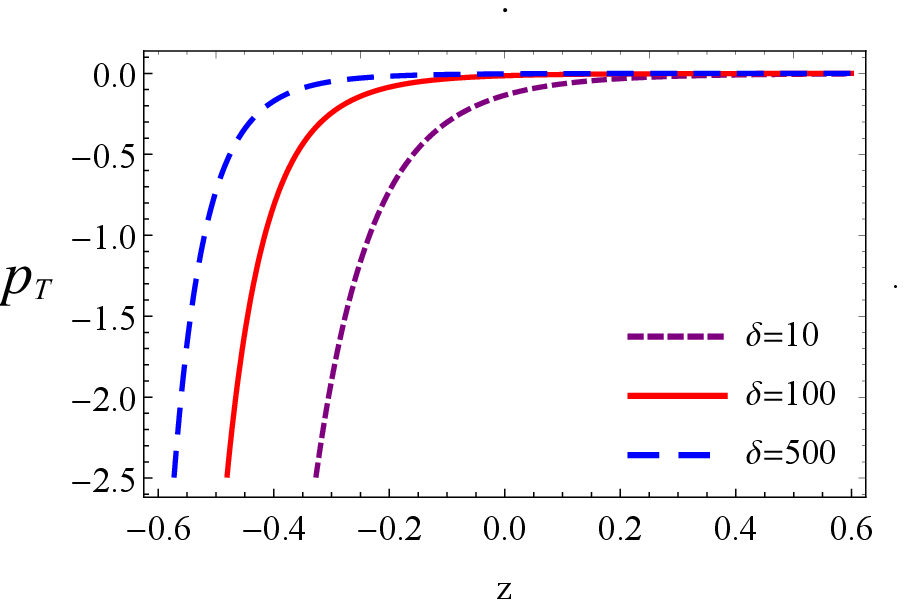}
   	\caption{a. Plot of  pressure $p_{_T} $  versus  $ z $  (H hoz.) cut-off b. Plot of pressure $ p_{_T} $ versus 
   	$z$  (GO hoz.) cut-off}
      \end{figure}
  %%%%%%%%%%%%%%%%%%%%%%%%%%%%%%%%%%%%%%%%%%%%%%%%%%%%%%%%%%%%%%%%%%%%%%%%%%%%%%%%%%%%%%%%%%%%%%%%%%%%%%%%%%%%%%%%%%%%%%%

  Figure $2a, b$ demonstrates the behaviour of pressure of RHDE with Hubble horizon and GO horizon cut-off versus redshift $z$ respectively. 
  Recent cosmological perceptions firmly show that our universe is dominated by the component with negative pressure called dark energy.\\
 
 It is ascertained that the isotropic pressure is negative throughout the evolution. From the figures we noticed that the pressure is decreasing 
 function of the redshift in both IR cut-offs.
 
  %%%%%%%%%%%%%%%%%%%%%%%%%%%%%%%%%%%%%%%%%%%%%%%%% Figures 3(a), 3(b), 3(c), 3(d) %%%%%%%%%%%%%%%%%%%%%%%%%%%%%%%%%%%%%%%%%%%%
 
 	\begin{figure}[H]
 	
 	\centering
 	(a)\includegraphics[width=7cm,height=7cm,angle=0]{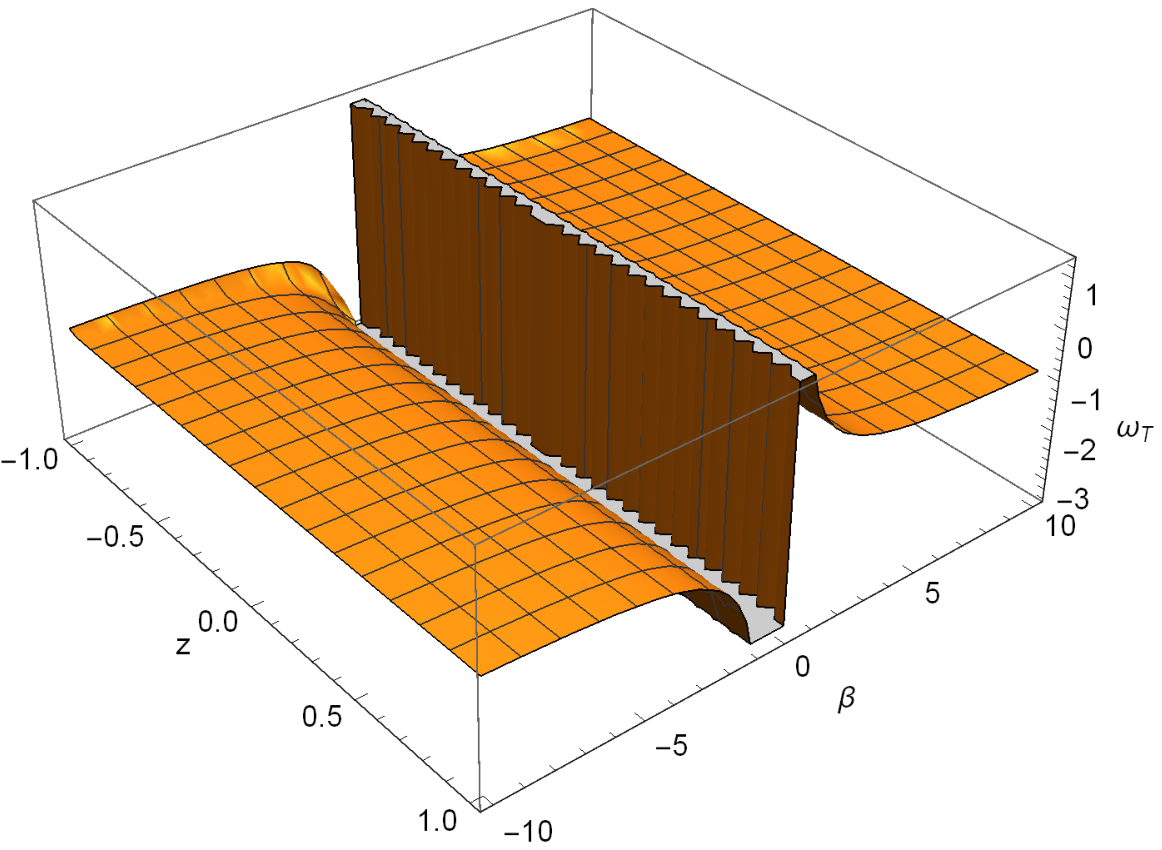}
 	(b)\includegraphics[width=7cm,height=7cm,angle=0]{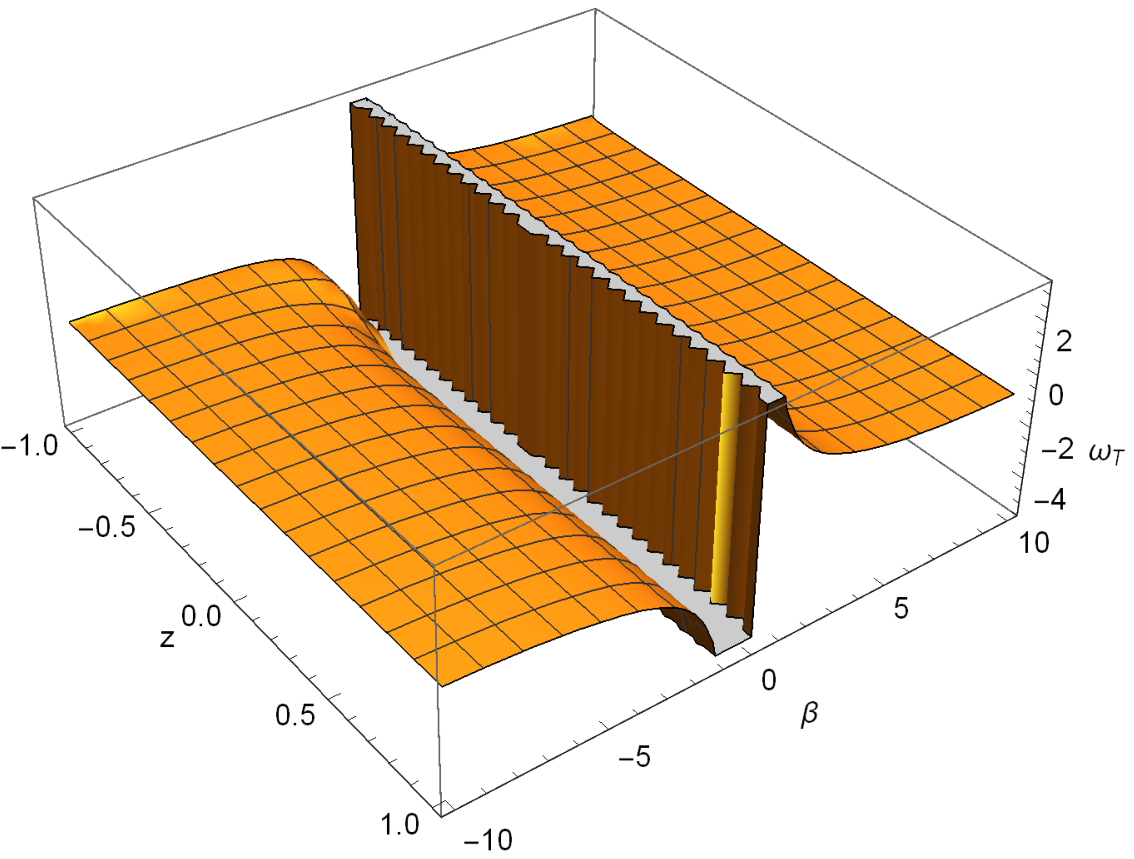}
 \caption{a. Plot of  EoS parameter $ \omega_{_T}$ versus $ z $ (H hoz.) cut-off  b. Plot of  EoS parameter 
 	$ \omega_{_T} $ versus  $ z $  (GO hoz.) cut-off}	
 \end{figure}
 %%%%%%%%%%%%%%%%%%%%%%%%%%%%%%%%%%%%%%%%%%%%%%%%%%%%%%%%%%%%%%%%%%%%%%%%%%%%%%%%%%%%%%%%%%%%%%%%%%%%%%%%%%%%%%%%%%%%%%%%%%%%%
 
 %%%%%%%%%%%%%%%%%%%%%%%%%%%%%%%%%%%%%%%%%%%%%%%%%%%% Table - 1 %%%%%%%%%%%%%%%%%%%%%%%%%%%%%%%%%%%%%%%%%%%%%%%%%%%%%%%%%%%%

\begin{table}[H]
	\caption{\small The behaviour of EoS parameter $\omega_{_T}$  with redshift parameterization for different values of  $\beta$ in 
	Hubble Horizon cut-off}
	\begin{center}
		\begin{tabular}{ c c|c c|c c }
			
			\hline
			\tiny & $\delta =10$  &\tiny & $\delta=100$ &\tiny &$\delta=500$  \\
			\hline
			\tiny	$\beta<0$	& \tiny Phantom	 &\tiny	$\beta<0$   & \tiny Phantom  & \tiny $\beta<0$& \tiny Phantom \\ 
			%\hline	

			\tiny	$0<\beta\le1.88$	& \tiny Quintessence	 &\tiny	$0<\beta\le1.98$   & \tiny Quintessence   & \tiny 
			$0<\beta\le1.99$ &\tiny Quintessence\\ 
			%\hline	
			
			\tiny	$1.88<\beta\le2$	& \tiny Quintessence$\to$ Phantom	 &\tiny	$1.98<\beta\le2$   & 
			\tiny  Quintessence$\to$ Phantom   & \tiny	$1.99<\beta\le10$ &\tiny  Quintessence$\to$ Phantom\\ 
			%\hline	
			\tiny	$2<\beta\le10$	& \tiny  Phantom	 &\tiny	$2<\beta\le10$   & \tiny   Phantom   & 
			\tiny	$2<\beta\le10$ &\tiny   Phantom\\ 
			%\hline	
			
			\end{tabular}
			\end{center}
			\end{table}
%%%%%%%%%%%%%%%%%%%%%%%%%%%%%%%%%%%%%%%%%%%%%%%%%%%%%%%%%%%%%%%%%%%%%%%%%%%%%%%%%%%%%%%%%%%%%%%%%%%%%%%%%%%%%%%%%%%%%%%%%%%%%%%%%%%
%%%%%%%%%%%%%%%%%%%%%%%%%%%%%%%%%%%%%%%%%%%%%%%%%%%%%% Table -2 %%%%%%%%%%%%%%%%%%%%%%%%%%%%%%%%%%%%%%%%%%%%%%%%%%%%%%%%%%%%%%%
\begin{table}[H]
	\caption{\small The behaviour of EoS parameter $\omega_{_T}$ with redshift parameterization for different values of  $\beta$ 
	in Granda--Oliver (GO) cut-off}
	\begin{center}
		\begin{tabular}{ c c|c c|c c }
			
			\hline
			\tiny & $\delta =10$  &\tiny & $\delta=100$ &\tiny &$\delta=500$  \\
			\hline
			\tiny	$\beta<0$	& \tiny Phantom	 &\tiny	$\beta<0$   & \tiny Phantom  & \tiny $\beta<0$& \tiny Phantom \\ 
			%\hline	

			\tiny	$0<\beta\le3$	& \tiny Quintessence	 &\tiny	$0<\beta\le3.7$   & \tiny Quintessence   & 
			\tiny	$0<\beta\le3.9$ &\tiny Quintessence\\ 
			%\hline	
			
			\tiny	$3<\beta\le3.99$	& \tiny quintessence$\to$ Phantom	 &\tiny	$3.7<\beta\le3.99$   & 
			\tiny  Quintessence$\to$ Phantom   & \tiny	$3.90<\beta\le4$ &\tiny  Quintessence$\to$ Phantom\\ 
			%\hline	
			\tiny	$4<\beta\le10$	& \tiny  Phantom	 &\tiny	$3.99<\beta\le10$   & \tiny   Phantom   & 
			\tiny	$4<\beta\le10$ &\tiny   Phantom\\ 
			%\hline	
			\end{tabular}
			\end{center}
			\end{table}
%%%%%%%%%%%%%%%%%%%%%%%%%%%%%%%%%%%%%%%%%%%%%%%%%%%%%%%%%%%%%%%%%%%%%%%%%%%%%%%%%%%%%%%%%%%%%%%%%%%%%%%%%%%%%%%%%%%%%%%%%%%%%%%%%%%%%

Figure $3a$ implies that the EoS parameter $\omega_{_T}$ is broadly used to categorize the different phases of the expanding universe. 
The EoS parameter of RHDE with $H$ hoz. cut-off  is given in Eq. (\ref{22}). The mathematical analysis of Eos parameter regarding 
quintessence / phantom for different values of $\beta$ in [--10, 10] has been presented in Table $1$ and $2$ respectively.
Here we notice that the EoS parameter initially lies  in low phantom region or phantom era $\omega_{_T}<-1$ at high redshift $z$ and 
crossing the phantom divide line(PDL) $\omega_{_T}=-1$. It enters  in quintessence region and again lies in phantom region for different 
values of $\delta$ are shown in Table $1$ and $2$. Similarly in Figure $3b$ additionally demonstrations the analysis of the EoS parameter for 
the (RHDE-GO cut-off) which is shown in Eq. (\ref{26}). In this figure we also observed that EoS parameter $\omega_{_T}$ moves towards the 
high phantom region to quintessence region and after that again lies in phantom region for different values of $\delta$. The trajectories of 
EoS parameter show the transition from phantom region to quintessence region by evolving the vacuum era of the universe in both IR cut-offs.
This is called quintom-like nature of the universe.

%%%%%%%%%%%%%%%%%%%%%%%%%%%%%%%%%%%%%%%%%%%%%%%%%%%%%%%%% Figures 4(a) & 4(b) %%%%%%%%%%%%%%%%%%%%%%%%%%%%%%%%%%%%%%

 \begin{figure}[H]
	
	\centering
	(a)\includegraphics[width=7cm,height=7cm,angle=0]{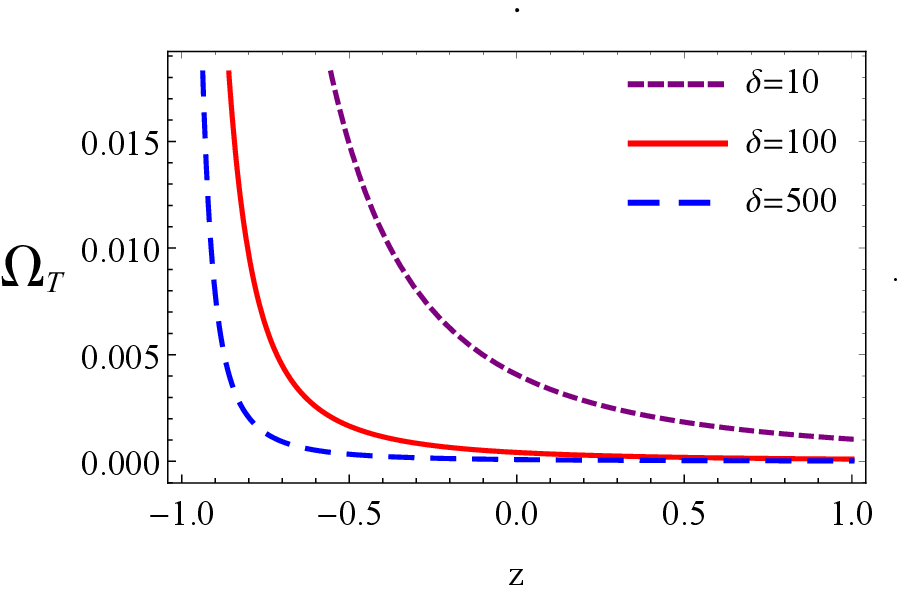}
	(b)\includegraphics[width=7cm,height=7cm,angle=0]{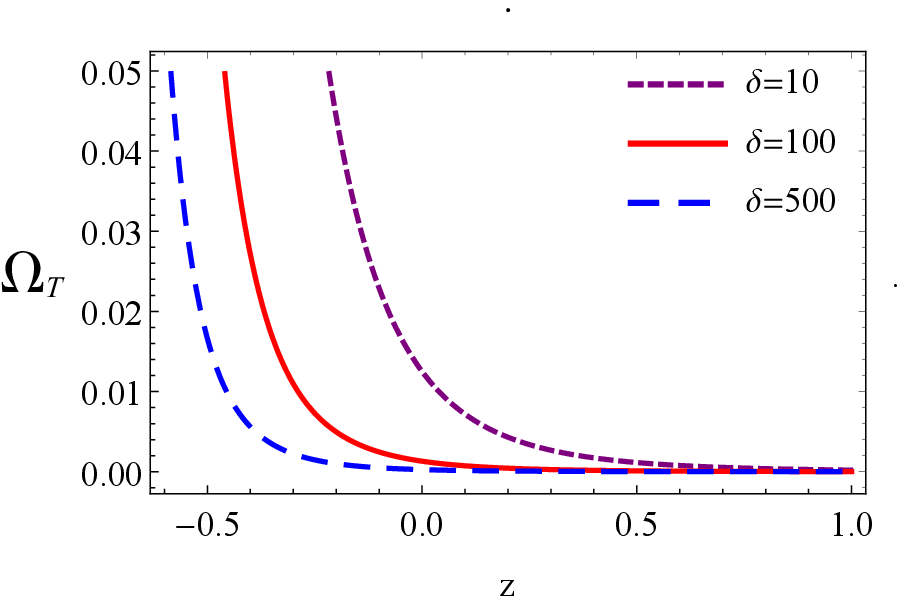}
	\caption{a. Plot of density parameter $ \Omega_{_T}$ versus $ z $  (H hoz.)  cut-off  b. Plot of density parameter $ \Omega_{_T}$ 
	versus  $ z $  (GO hoz.) cut-off.}	
\end{figure}
%%%%%%%%%%%%%%%%%%%%%%%%%%%%%%%%%%%%%%%%%%%%%%%%%%%%%%%%%%%%%%%%%%%%%%%%%%%%%%%%%%%%%%%%%%%%%%%%%%%%%%%%%%%%%%%%%%%

Fig. $ 4$ depicts the behaviour of density parameter $\Omega_{_T}$ versus redshift $z$ for various estimations of $\delta$.
In this figure the density parameter of the RHDE is seen to be lower region for the high redshift. We also noticed  that the density parameter 
is positive throughout the evolution in both IR cut-offs.
 
%%%%%%%%%%%%%%%%%%%%%%%%%%%%%%%%%%%%%%%%%%%%%%%%% Section 4 %%%%%%%%%%%%%%%%%%%%%%%%%%%%%%%%%%%%%%%%%%%%%%%%%%%%%%%%%%%%%%%%

\section{Dynamics of $\omega_{_T}$-$\omega_{_T}^{'}$ plane}

Caldwell and Linder \cite{ref81,ref82}, recommend the $\omega_{_T}$-$\omega_{_T}^{'}$ plane to describe the dynamical property of DE model. 
Here  $\omega_{_T}$ is the EoS parameter and $\omega_{_T}^{'}$  is its evolutionary structure. Where dot means the derivative concerning $ln a$.
Here the ($\omega_ {_T}$-$\omega_{_T}^{'}$) plane was segregated in two regions, the thawing region $\omega_{_T}<0, $ $\omega_{_T}^{'}>0 $ is 
the region where the EoS parameter evolves $\omega_{_T}<-1 $ increases with time and its evolution parameter shows positive behaviour, and 
the freezing region $\omega_ {_T}<0, $ $\omega_ {_T}^{'}<0 $ in this region evolution parameter is negative.\\

For R$\grave{e}$nyi HDE with (H hoz.) IR-cut-off with redshift\\

We can obtain the $\omega_{_T}^{'}$ by differentiating the EoS parameter which is in Eqs. (\ref{22}) and (\ref{26}) with respect to $lna$ then we get
\begin{equation}
\label{29}
\omega'_{_T}=\frac{d \omega_{_T}}{d \ln a}= \frac{108 \pi  \delta  \left(\frac{a_0}{z+1}\right){}^{3/\beta }}
{\left(9 \pi  \delta  \left(\frac{a_0}{z+1}\right){}^{3/\beta }+4 \beta ^2\right){}^2}.
\end{equation}

\  For R$\grave{e}$nyi HDE with (GO hoz) IR cut-off  with redshift\\
\begin{equation}
\label{30}
\omega'_{_T}=\frac{d \omega_{_T}}{d \ln a}= \frac{3888 \pi  \delta  \left(\frac{a_0}{z+1}\right){}^{6/\beta } (3 m-2 \beta  k)^2}
{\left(81 \pi  \delta  \left(\frac{a_0}{z+1}\right){}^{6/\beta }+16 \beta ^4 k^2-48 \beta ^3 k m+36 \beta ^2 m^2\right){}^2}.
\end{equation}

%%%%%%%%%%%%%%%%%%%%%%%%%%%%%%%%%%%%%%%%%%%%%%%%%%%%%%%%%%%%%%%% Figures 5(a) & 5(b) %%%%%%%%%%%%%%%%%%%%%%%%%%%%%%%%%%%%%%%%%%%%
\begin{figure}[H]
	\centering
	(a)\includegraphics[width=7cm,height=7cm,angle=0]{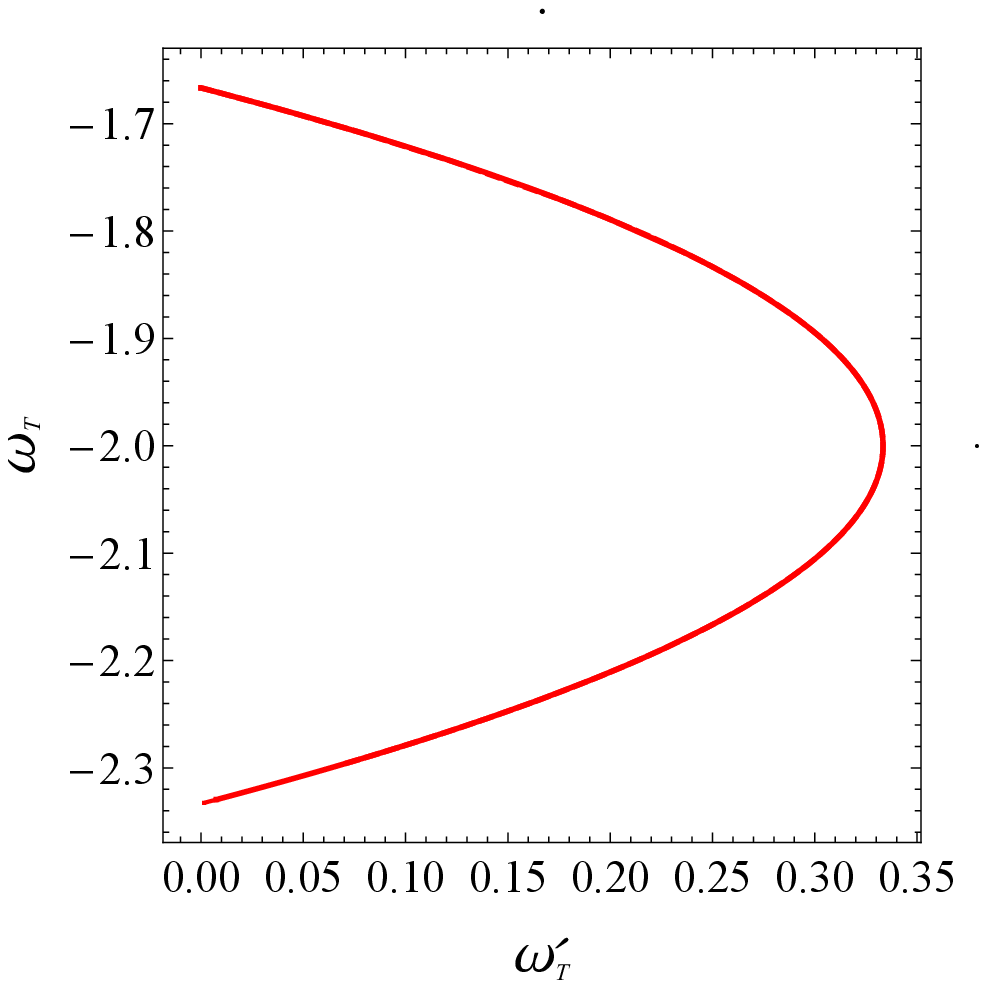}
	(b)\includegraphics[width=7cm,height=7cm,angle=0]{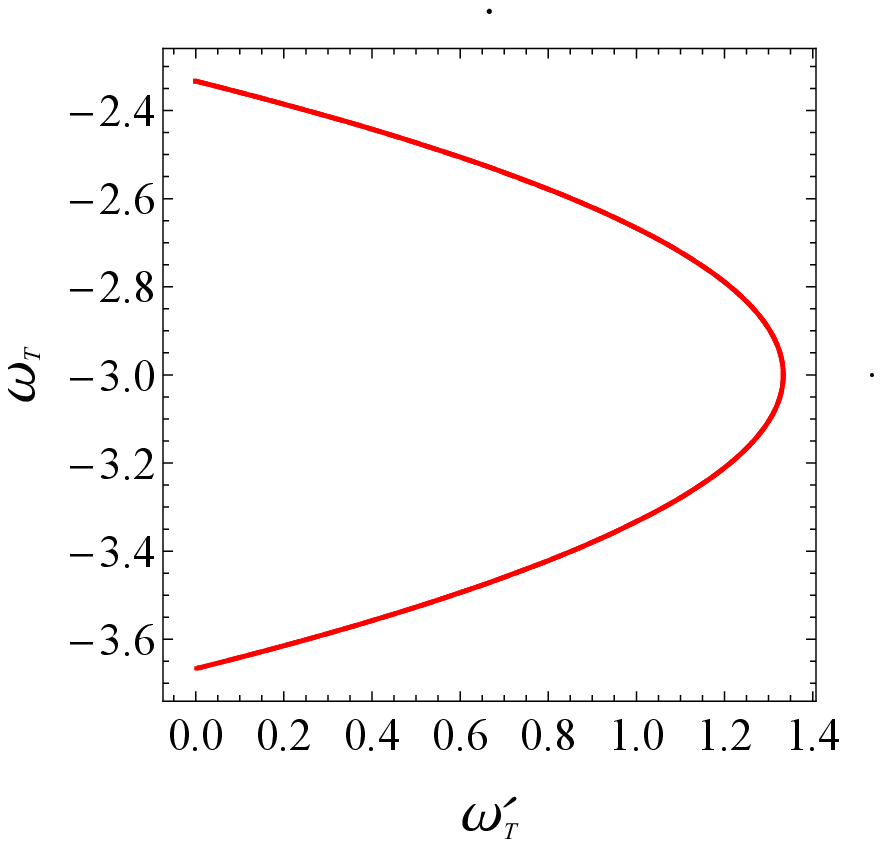}
	\caption{a. Plot of $ \omega_{_T}-\omega_{_T}^{'}$ versus $ z $ (H hoz) cut-off  b. Plot of  $ \omega_{_T}-\omega_{_T}^{'}$ versus  
	$ z $  (GO hoz.) cut-off}	
\end{figure}
%%%%%%%%%%%%%%%%%%%%%%%%%%%%%%%%%%%%%%%%%%%%%%%%%%%%%%%%%%%%%%%%%%%%%%%%%%%%%%%%%%%%%%%%%%%%%%%%%%%%%%%%%%%%%%%%%%%%%%%%%%%%

From figure $5a, b$ we discuss the behaviour of ($\omega_{_T}$-$\omega_{_T}^{'}$) plane corresponding to 
Eqs. (\ref{29}) and (\ref{30}) for three different values of $\delta$. These figures  shows that ($\omega_{_T}$-$\omega_{_T}^{'}$) 
trajectories represents thawing region and  are overlapped together  during the whole evaluation. 

%%%%%%%%%%%%%%%%%%%%%%%%%%%%%%%%%%%%%%%%%%%%%%%%%%%% Section 5 %%%%%%%%%%%%%%%%%%%%%%%%%%%%%%%%%%%%%%%%%%%%%%%%%%5

\section{Stability of the model}

One may check the stability of the derived solution with respect to the perturbation
of the space-time \cite{ref83}-\cite{ref84}. To this purpose, we considered the perturbations of volume scalar, directional Hubble factors and 
mean Hubble factor as

\begin{equation}
\label{31}
V \rightarrow V_{B} + V_{B} \sum_{i=1}^3 \delta b_{i},~~~
H_{i} \rightarrow H_{B_{i}} + \delta \dot b_{i}, ~~~
H\rightarrow H_{B}+\frac{1}{3}\sum_{i=1}^3 \delta \dot b_{i}, ~~~
\sum_{i=1}^3 H_{i}^{2} \rightarrow \sum_{i=1}^3 H_{{B}_i}^{2}+2 \sum_{i=1}^3 H_{{B}_i}. \delta b_{i}. 
\end{equation}

For metric perturbation $ \delta b_i $ to be linear the following equations must be satisfied
\begin{equation}
\label{32}
\sum_{i}{\ddot{\delta b_{i}}}+2 \sum_{i} \theta_{B_{i}} \dot{\delta b_{i}}=0,
\end{equation}
\begin{equation}
\label{33}
\ddot{\delta b_{i}}+\frac{\dot{V_{B}}}{V_{B}}\dot{\delta b_{i}}+\sum_{j} \dot{\delta b_{j}}\theta_{B_{i}}=0,
\end{equation}
\begin{equation}
\label{34}
\sum \dot{\delta b_{i}}=0.
\end{equation}
From  the simplification of above equations 

\begin{equation}
\label{35}
\ddot{\delta b_{i}}+\frac{\dot{V_B}}{V_B}\dot{\delta b_i}=0,
\end{equation}
where the background volume scalar $ V_B $ leads to 
\begin{equation}
\label{36}
V_B=t^{2 \beta}.
\end{equation}

From Eqs.(\ref{35}) and (\ref{36}), the metric perturbation becomes
\begin{equation}
\label{37}
\delta b_i=c_2+c_1 \frac{t^{1-2 \beta }}{(1-2 \beta) },
\end{equation}
where $ c_1 $ and $ c_2 $ are constants of integration.\\

Thus, the actual fluctuation for each expansion factor $ \delta a_i=a_{B_{i}}\delta b_i $ are expressed as
\begin{equation}
\label{38}
\delta a_i= c_2 t^{2 \beta }+c_1 \frac{t}{(1-2 \beta) }.
\end{equation}

%%%%%%%%%%%%%%%%%%%%%%%%%%%%%%%%%%%%%%%%%%%%%%%%%%%%%%%% Figure 6 %%%%%%%%%%%%%%%%%%%%%%%%%%%%%%%%%%%%%%%%%%%%%%%%%%%%%

\begin{figure}[H]
\centering
\includegraphics[width=9cm,height=9cm,angle=0]{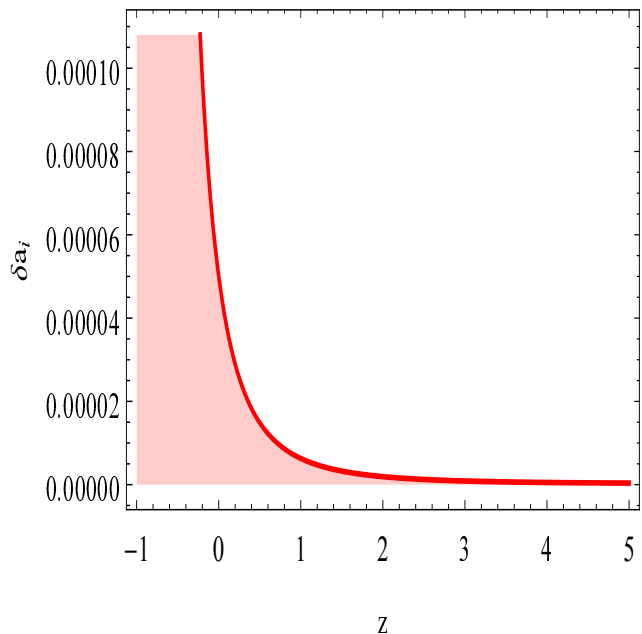}
\caption{ Plot of stability analysis with redshift $z$}
\end{figure}
%%%%%%%%%%%%%%%%%%%%%%%%%%%%%%%%%%%%%%%%%%%%%%%%%%%%%%%%%%%%%%%%%%%%%%%%%%%%%%%%%%%%%%%%%%%%%%%%%%%%%%%%%%%%%%%%%%%

From Eq.(\ref{38}) and  Fig. $6$, we notice that the positive estimation of $\beta$, the $\delta_{ai}$ approaches to zero for 
large $z\to\infty$ i.e. $\delta_{ai}\to 0$. Thus, the background solution is stable against the metric perturbation.
This figure depicts the behaviour of actual fluctuations with respect to redshift. In past, the actual fluctuation is null but it 
increases minimally on later time.

%%%%%%%%%%%%%%%%%%%%%%%%%%%%%%%%%%%%%%%%%%%%%%%%%%%%%%%%% Section 6 %%%%%%%%%%%%%%%%%%%%%%%%%%%%%%%%%%%%%%%%%%%%%%%%%

\section{Conclusion} 

 In this work, the R$\grave{e}$nyi holographic type DE model with two IR cut-offs: Hubble horizon $ L=\frac{1}{H} $ and Granda \& Oliveros horizon  
 $ L=(k H^2+m \dot{H})^{-1/2}$ cut-off have been discussed in $f(R, T)$ gravity by considering the parameters  $k= 0.8$  and $m = 0.89$. 
 For this purpose , we have analysed the cosmological features like: energy density, pressure, EoS parameter, density parameter, 
 ($\omega_{T}$-$\omega_{T}^{'}$) plane and stability of the models are discussed. We have summarized our results as follows: \\
 
\begin{itemize}

 \item The energy density of RHDE $\rho_{_T}$ is positive decreasing function for the three values of $\delta$ with redshift $z$. We found 
 this behaviour is similar in both IR cut-offs as seen in Fig.$1a, b$ respectively.
 
 \item Both models (I, II) indicate that the pressure is negative in all estimations throughout the evaluation seen in Fig.$2a, b$.
 
 \item The trajectories of EoS parameter show the transition from phantom region to quintessence region by evolving the vacuum era of the 
 universe in both IR cut-offs as clear from Fig.$3a, b$.

 \item We also notice that the density parameter $\Omega_{_T}$ is positive throughout the evolution in both IR cut-offs as seen in Fig.$4a, b$.

 \item The ($\omega_ {_T}$-$\omega_ {_T}{'}$) trajectories indicate the thawing region for the R$\grave{e}$nyi HDE model with Hubble horizon 
 cut-off and GO cut-off as seen in Fig.$5a, b$.

 \item We have conducted that the perturbation analysis to examine the stability of the considered dark energy model and found that RHDE 
 shows more stability during the comic evolution seen in (Fig.$6$). Subsequently our created model is stable.
 
 \item Here, we also include a comparative study of our work with recent works on this subject\cite{ref34,ref35} and found that our findings are 
 in good agreement with their work but their models are unstable whereas our both models are stable.  

\end{itemize}

The feasibility of the R$\grave{e}$nyi holographic dark energy model with Hubble and GO cut-off is supported by our research and combined 
observational data is consistent with the future horizon.
%%%%%%%%%%%%%%%%%%%%%%%%%%%%%%%%%%%%%%%%%%%%%%%%%%%%%%%%%%%%%%%%%%%%%%%%%%%%%%%%%%%%%%%%%%%%%%%%%%%%%%%%%%%%%%%%%%%%%%%%%%%%%%%%%%%%
\section*{Acknowledgments} 
The authors are heartily grateful to the anonymous referee for his constructive comments which improved
the paper in the present form.
%%%%%%%%%%%%%%%%%%%%%%%%%%%%%%%%%%%%%%%%%%%%%%%%%%%%%%%%%%%%%%%%%%%%%%%%%%%%%%%%%%%%%%%%%%%%%%%%%%%%%%%%%%%%%%%%

\end{document}